# Nanoscale structural characterization of manganite thin films integrated to silicon correlated with their magnetic and electric properties.

Aneely Carrero [a, b, c], Augusto Roman [a, b, c], Myriam Aguirre [d, e, f], Laura B. Steren [a, b, c]

[a] Instituto de Nanociencia y Nanotecnología, B1650 San Martín, Buenos Aires, Argentina.
[b] Consejo Nacional de Investigaciones Científicas y Técnicas, C1425FQB Ciudad Autónoma de Buenos Aires, Argentina.
[c] Laboratorio de Nanoestructuras Magnéticas y Dispositivos, Dpto. Materia Condensada, Centro Atómico Constituyentes, B1650 San Martín, Buenos Aires, Argentina
[d] Departamento de Física de la Materia Condensada, Universidad de Zaragoza, E-50009 Zaragoza, Spain.
[e] Instituto Universitario de Nanociencia Aragón e Instituto de Ciencia de Materiales de Aragón-CSIC, E-50018 Zaragoza, Spain.
[f] Laboratorio de Microscopías Avanzadas, Universidad de Zaragoza, E-50018 Zaragoza, Spain.

A detailed nanoscale structural characterization was performed on high-quality $La_{0.66}Sr_{0.33}MnO_3$ (LSMO) thin films of different thicknesses and deposited by pulsed laser deposition onto buffered Si (100) substrates. A multilayered structure built of $Y_{0.13}Zr_{0.87}O_2$ (YSZ) and $CeO_2$ layers was used as buffer in order to optimize the manganite films growth. The stacking of the different layers, their morpholohy, composition and strains were analysed using different experimental techniques. In-situ characterization of the films, performed with reflection high-energy electron diffraction, revealed their epitaxial growth and smooth surfaces. High-resolution transmission electron microscopy (HR-TEM) images showed sharp interfaces between the constituents lattices and combined with energy-dispersive X-ray analysis allowed us to determine that there was no ion interdifussion across them. The Fourier-Fast-Transform of the HR-TEM images was used to resolve the epitaxy relationship between the layers, resulting in [100] LSMO (001) ∥ [110] $CeO_2$ (001) ∥ [110] YSZ (001) ∥ [110] Si (001). The LSMO thin films were found to be ferromagnetic and metallic at low temperature regardless their thickness. The effect of strains and defects was only detected in films thinner than 15 nm and put in evidence by X-ray diffraction patterns and correlated with magnetic and electrical parameters.

## 1. Introduction

During the last years, there has been an increasing interest in the integration of complex oxides with silicon wafers looking for original spintronic devices, taking advantage of their rich variety of physical properties [1, 2]. The cooperative effect of spin, orbital and charge degrees of freedom in one of the more important variety of mixed oxides, i.e. the perovskites $ABO_3$, leads to multiferroic orders attractive not only for fundamental science research but for potential applications [3, 4]. The electronic and magnetic properties of perovskites are extremely sensitive to the octahedral environment of the 3d ions in B position and any distortion of the oxygen octahedra, due to the oxide composition - i.e. A cations size [5] - or strains [6] affect the properties of these materials. There are many reports on different strategies developed to tackle the growth of perovskites films onto silicon. A large variety of deposition techniques [7-10] and buffers [11-14] have been employed, in particular, to attain epitaxial $La_{0.67}Sr_{0.33}MnO_3$ (LSMO) ferromagnetic films on silicon substrates minimizing the effect of faults and strains. LSMO is a half-metal ferromagnet, already tested as metallic electrode in magnetic tunnel junctions [15] and combined with ferroelectric in artificial multiferroic heterostructures [16]. In both examples, the LSMO layers were deposited onto oxide substrates. The growth of LSMO onto Si has shown that, parameters such as the Curie temperature and the saturation magnetization of ferromagnetic LSMO together with the magneto-transport properties are extremely sensitive to buffer layers and growth techniques. It is crucial, thus, to characterize the stacking and the crystalline structure of oxides heterostructures and to correlate them to their physical properties. In this frame, the growth of epitaxial oxides structures onto silicon wafers is still a challenge, owing to the high reactivity of silicon surface to oxygen and the cationic interdiffusion but mainly due to the mismatch of thermal expansion coefficients and lattices between both materials [14]. Some of these aspects can be overcome by intercalating buffer layers between the substrate and the perovskite.

In this paper, high-quality epitaxial ferromagnetic manganites $La_{0.67}Sr_{0.33}MnO_3$ (LSMO) thin films were grown by Pulsed Laser Deposition (PLD) onto buffered silicon substrates. A detailed study of LSMO films deposited onto $CeO_2$/ $Y_{0.13}Zr_{0.87}O_2$ //Si (100) was performed. Yttria-Stabilized Zirconia $Y_{0.13}Zr_{0.87}O_2$ (YSZ) is generally used as buffer onto silicon substrates, since it decomposes the $SiO_2$ of the Si surface by forming $ZrO_2$ and desorbing $SiO_x$ . [17] YSZ results in a effective barrier to oxygen interdiffusion. LSMO deposited directly onto YSZ//Si grows polycrystalline, so it is necessary to introduce a multilayer buffer to get epitaxial structures. [18] It is well-known that the lattice mismatch between the film and the substrate or the closest layer in the case of multilayers, $\zeta$, should be small enough to favour epitaxial growth being $\zeta = ((a_b - a_u)/a_u) \times 100$ where $a_b$ and $a_u$ are the lattice parameters of the bottom and the upper layer respectively. [19] We chose $CeO_2$ as an intercalating layer in order to get a better matching of the LSMO layers with the buffered substrates. An exhaustive structural and morphological characterization of the samples was carried out to analyze the magnetic and transport properties of the samples. The films' growth was monitored by Reflected High Energy Electrons Diffraction (RHEED) while the out-of-plane lattice parameters were determined by X-ray diffraction (XRD) experiments. High-resolution transmission electron (TEM) and atomic force (AFM) microscopies were used to examine the quality of surfaces and interfaces, the stacking and composition of the differente constituents. Finally, the magnetic and transport properties of the samples were analysed and correlated with structural properties.

## 2. Experimental details

A series of $LSMO/CeO_2/YSZ$//Si(100) samples was grown for this study by Pulsed Laser Deposition using a Nd: Yag laser operating at 266 nm with a fluency of about 2.0 J/cm$^2$. A multi-target stage with ceramic YSZ, $CeO_2$ and LSMO was used to fabricate the multilayers in a single process. The growth of the multilayers was optimized by varying the chamber pressure, the substrate temperature and the target to substrate distance in order to get epitaxial structures. Epitaxial films were finally obtained keeping the substrate temperature at 800 °C and changing the chamber pressure while growing the different constituents of the structures, i.e. YSZ at $P_{O2}$= 0.04 Pa, $CeO_2$ at $P_{O2}$= 0.05 Pa and LSMO at $P_{O2}$= 30 Pa. The growth rates for these layers were 0.16 nm/sec, 0.90 nm/sec and 2.7 nm/sec, respectively. After deposition, the samples were cooled down to room temperature under an oxygen pressure of 40 kPa. the growth conditions of the buffer layers were kept fixed. The LSMO layers thickness was varied from 10 nm to 80 nm. A LSMO thin film deposited on (100) STO was grown in the same conditions to use as reference.

The multilayers were characterized by multiple techniques. The growth of the films was monitored using in-situ RHEED. The stacking, interfaces and crystallinity of the different layers were analysed by scanning TEM coupled with a high angle annular dark field detector in a FEI Titan G2 at 300 keV probe corrected (a CESCOR Cs-probe corrector from CEOS Company) and fitted with an energy dispersive X-ray (EDX) spectrometer system by EDAX. The samples for TEM were prepared by Focused Ion Beam (FIB) by Helios 650, with profile of reduced thickness of 50 nm thick fort he electron transparency. X-ray diffractograms were obtained by a Panalytical- Empyrean diffractometer with detector PIXcel3D and radiation CuKα in the parallel-beam geometry. The morphology of the films´surfaces was examined using a Bruker Atomic Force Microscope. The magnetic measurements were carried out in a commercial superconducting quantum interference device magnetometer. Hysteresis loops and magnetization vs. temperature curves were measured between 5 K – 300 K for fields up to 5 T. The magneto-transport measurements were performed using the four-probe method, with the magnetic field applied in the plane of the multilayers.

## 3. Results and discussion
### 3.1. Stacking, crystalline structure and morphology

The growth of the different layers into the LSMO/$CeO_2$/YSZ//Si structure was analysed by in-situ RHEED. The typical RHEED patterns obtained for the different components of the multilayers are plotted in Fig. 1. The streaky patterns of the buffer layers (Fig. 1 (a-d)) suggest a growth with small domains and atomically flat surfaces [20]. The patterns of the LSMO display, instead, weakly modulated streaks (Fig.1 (e-f)) associated to smooth surfaces with stepped surface [21]. Roughness, measured by AFM, reveals that the films surfaces are, in fact, very smooth (data not shown). The root mean square roughness (RMS) of the buffers, calculated in a 1x1 μm area, is about 0.5 nm and increases to about 1nm for the LSMO films. AFM results agreed with the analysis of the RHEED patterns. The roughness of our films was comparable to those obtained for films grown by molecular beam epitaxy (MBE) [9] and combined polymer-assisted-deposition/MBE [10] onto $SrTiO_3$ buffers. The wavy profile of the LSMO surface, observed in TEM images, results in a 2D granulated surface of approximately 1 nm height and 32 nm mean diameter, computed in a 1x1 μm$^2$ area scan.

The crystalline structure of the multilayers was studied through XRD in the θ-2θ configuration. A typical XRD pattern of a LSMO/$CeO_2$/YSZ//Si structure is shown in Fig.2. The spectra only present the (00h) reflections of the LSMO, YSZ and $CeO_2$ layers. No additional peaks of other crystal orientations or phases were detected in any sample. A zoom of the (002) LSMO Bragg reflections for two different samples (inset of Fig. 2) reveals that the position and intensity of this peak depend notably with the LSMO thickness. On one hand, the (002) peak shifts to lower angles as the film thickness decreases. This result indicates an increase of the out-of plane lattice parameter as a consequence of in-plane compressive strains of the lattice as it is adjusted to the $CeO_2$ cell. On the other hand, the reflection broadening observed as the films thickness decreases is associated to strains and lattice faults. These effects are expected to be more remarkable in thinner films [22].

In Fig. 3 (a) a TEM image of the LSMO/CeO$_2$/YSZ//Si heterostructure is shown and its constituents labeled. All the interfaces look sharp while the thickness of each layer is uniform, and were determined to be 75 nm, 30 nm and 20 nm for the LSMO, CeO$_2$ and YSZ layers, respectively. The manganite films have usually wavy surfaces, independently of their thickness. This morphology is usually observed in perovskite oxides grown on silicon by PLD and it is probably a consequence of strains induced along the stacking of the heterostructure [23, 24]. The analysis of the interfaces was made by high resolution TEM (HR-TEM) - Figs. 3 (b - d). Typically the YSZ//Si interface, shown in Fig. 3 (b), reveals the existence of a 3 nm thick amorphous SiOx layer. The roughness at this interface is around 1 nm. The YSZ layer is epitaxial and has a well-defined interface with the CeO$_2$ layer (Fig. 3 (c)). Finally, HR-TEM images of the LSMO/CeO$_2$ interfaces (Fig. 3 (d)) show a roughness close to 1 nm. The epitaxial growth of the LSMO is also appreciated from this image.

The strain state of the multilayers was investigated by General Phase Analysis (GPA) on High-Angle Annular Dark-Field scanning TEM images [25, 26]. Typical color-coded two-dimensional out-of-plane (Eyy) and in-plane (Exx) strain maps are shown in Figs. 4 (b-c). According to the color scale [27, 28, 29], and in spite of the high degree of defects, the maps suggest that YSZ is compressed in the plane (x) of the films and slightly expanded along the y-direction. The high degree of defects in the case of CeO$_2$ prevents to do the same analisis for this layer. The LSMO film instead is notably relaxed with few defects at the CeO$_2$/LSMO interface. The GPA was performed taken the unstrained silicon substrate as reference.

The Fast Fourier Transform (FFT) of the HR-TEM images was used to evaluate the epitaxial relationship between layers. Fig. 5(a-d) shows the electron diffraction patterns of the different layers, resulting in the following epitaxial relationship [100] LSMO (001) ∥ [110] CeO$_2$ (001) ∥ [110] YSZ (001) ∥ [110] Si (001).—The 45º rotation of the LSMO perovskite cell in the plane of the films reduces the elastic energy by optimizing the accomodation of the smaller LSMO lattice onto the CeO$_2$ one (Fig. 5 (e,f)). As a consequence, the misfit parameter between LSMO and CeO$_2$ reduces from 40% to less than 1%. A similar rotation was observed by M. Scigaj and coworkers [30] when depositing YSZ onto SrTiO$_3$.

The composition profile of the samples was probed along the [001] direction by EDX (Fig. 6). The detection of the constituents of each layer is clear and so are the well-resolved interfaces. Regarding the multilayer composition, YSZ and CeO$_2$ were found to preserve their stoichiometry, while the LSMO film is slightly oxygen-deficient and free of impurities.

**3.2. Magnetism and electric transport**

The magnetic properties of the samples were examined by magnetization measurements as a function of magnetic field and temperature. The temperature dependence of the magnetization (Fig. 7(a)) shows a paramagnetic-ferromagnetic transition around 350K, saturating at around Ms=5.7x10$^5$ A/m for the thicker film.
The Curie temperature, Tc, was determined as the temperature at which a sudden rise of magnetization is observed. These values are very close to the bulk values [31] indicating that the thick LSMO layers are almost fully relaxed and stoichiometric in agreement with structural and compositional results. Thinner films exhibit lower Tc, featuring the effect of strains onto the first nm of the samples. [32] Magnetization loops of the LSMO/CeO$_2$/YSZ//Si (100) structures are shown in Fig. 7(b). The ferromagnetic hysteresis are sistematically larger than those measured for LSMO/SrTiO$_3$, although the thicker films have similar values of coercivity, Hc and remanence over saturation magnetizatin ratio, Mr/Ms (Fig. 7 (c)). We observed that Hc decreases while Mr/Ms slightly increases as the buffered LSMO thickness increases. [33] This behaviour is explained by a major effect of strains and presence of lattice faults and distortions in the region close to the LSMO/$CeO_2$ interface, according to what was deduced from XRD patterns and HR-TEM images.

The temperature dependence of the resistivity of 75 nm-thick and 10 nm-thick LSMO buffered films are shown in Figs. 8(a, b). As observed, both samples exhibit a metal-insulator transition (MIT). Thick samples show a MIT close to their Curie temperatures. The thinner LSMO layer, on the other hand, exhibit a broad rounded MIT well-below the magnetic transition. Furthermore, the resistivity of the thinner sample is an order of magnitude higher thant that of the thick film. Once more, these results put in evidence the improvement of the quality of the films as their thickness is increased, in accordance with the structural and magnetic properties described above. A Colossal Magnetoresistance effect is measured for both samples around the MIT (not shown) as expected for LSMO compounds. [34]

**CONCLUSIONS**

We have successfully grown high-quality LSMO thin films onto buffered silicon by PLD, performing a detailed study of their structural properties at the nanometer scale, including analysis of stacking, composition and strains. The manganites layers have been proved to exhibit excellent crystallinity and flat surfaces using different experimental techniques. We have found that multilayer interfaces are sharp, not only from a structural point of view but also with respect ions interdiffusion. We´ve also shown that the LSMO cells are rotated with respect to the underlayer ones, so as to minimize the lattice mismatch between both compounds. By optimizing the growth parameters and choosing appropriate buffers we´ve obtained thin films with excellent ferromagnetic and metallic properties, close to that of bulk compounds. The effect of defects, stacking faults and roughness in the properties of the films was carefully analyzed not only by their structural characterization but also by their magnetic parameters and their electrical resistivities.

**ACKNOWLEDGEMENTS**

The TEM microscopy studies have been conducted at the "Laboratorio de Microscopias Avanzadas" of the "Instituto de Nanociencia de Aragón - Universidad de Zaragoza". Authors acknowledge the institution for offering access to their instruments and expertise. The authors wish to thank H. Corti and his team at the INN for the use of the AFM. This work was supported by FONCYT PICT 2014-1047 and 2016-0867, CONICET PIP 112-201501-00213, MINCYT and the H2020-MSCA-RISE-2016 SPICOLOST (N° 734187).

**Figure captions**

Fig. 1 RHEED patterns with electron beam incident along [110] and [100] directions of (a, b) YSZ, (c, d) $CeO_2$ and (e, f) LSMO, respectively.

Fig. 2 XRD pattern in the θ-2θ configuration of the 75 nm thick LSMO films. Inset: Detail of the XRD patterns around LSMO (200) Bragg reflection for a 25 nm (blue) and a 75 nm (pink) thick films.

Fig. 3 (a) TEM cross-sectional image of the LSMO 75 nm/$CeO_2$/YSZ//Si heterostructure. Detail of HR-TEM images at (b) YSZ/Si (c) $CeO_2$/YSZ and (d) LSMO/$CeO_2$ interfaces.

Fig.4 (a) Cross-sectional HR-TEM image of a LSMO/$CeO_2$/YSZ//Si multilayer. In-plane, Exx, and out-of-plane, Eyy, strain maps of the heterostructure calculated by GPA are shown in (b) and (c), respectively.

Fig. 5 (a-d) Electron diffraction patterns from (a) Si, (b) YSZ, (c) $CeO_2$, (d) LSMO, respectively. (e,f) Schematic cartoons of the CeO2 and LSMO crystal structures matching, according to [100] LSMO (001) ∥ [110] $CeO_2$ (001) epitaxial relationship. (e) is a lateral view and (f) a top view.

Fig. 6 EDX linescan profile of the LSMO 75 nm/$CeO_2$/YSZ//Si multilayer.

Fig.7 (a) Magnetization vs. temperature measured with an applied magnetic field of 0.5 T and (b) Magnetization vs. magnetic field curves for a (□) LSMO 75 nm and (●) LSMO 10 nm buffered structures, measured at 10 K. (c) (●) Coercive field, Hc and (■) remanence over saturation magnetization ratio, Mr/Ms as a function of the films thickness for LSMO buffered structures. The dashed lines are a guide to the eyes.

Fig.8 Resistivity vs. temperature of the (a) 75 nm and (b) 10 nm - thick LSMO films integrated onto buffered silicon.

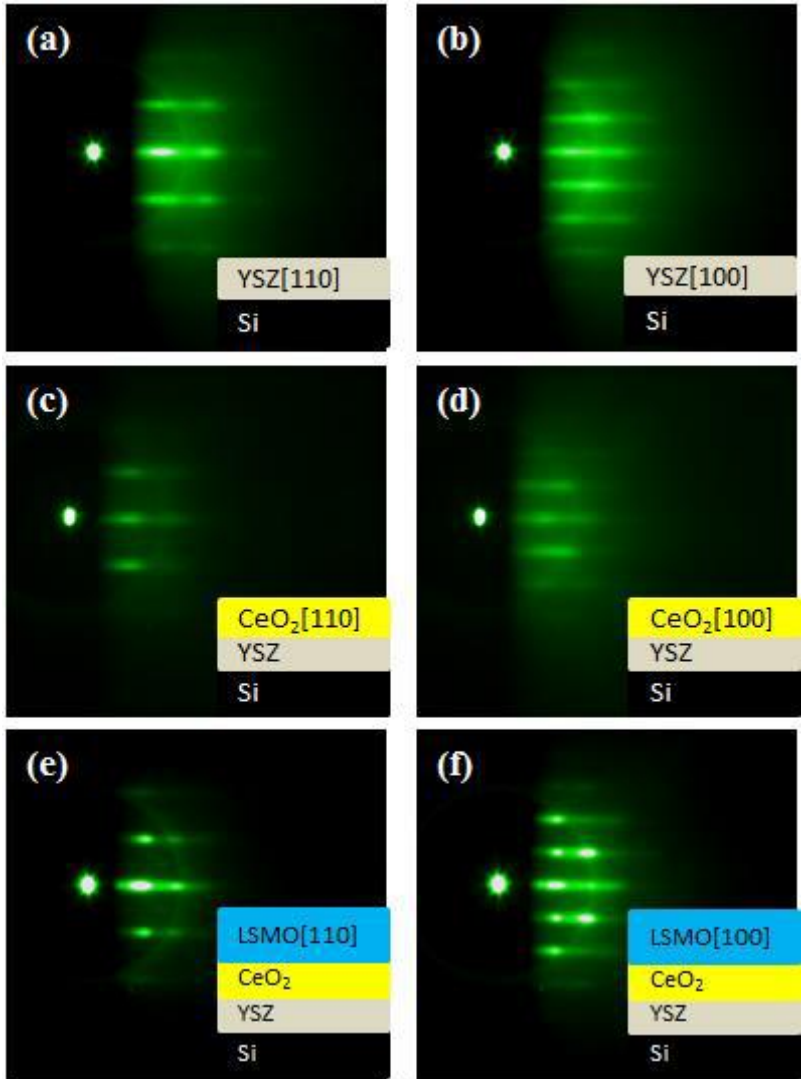

Fig. 1

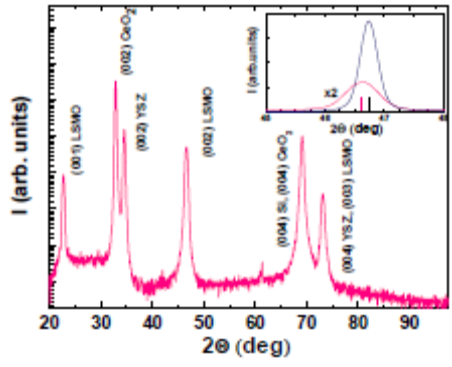

**Fig. 2**

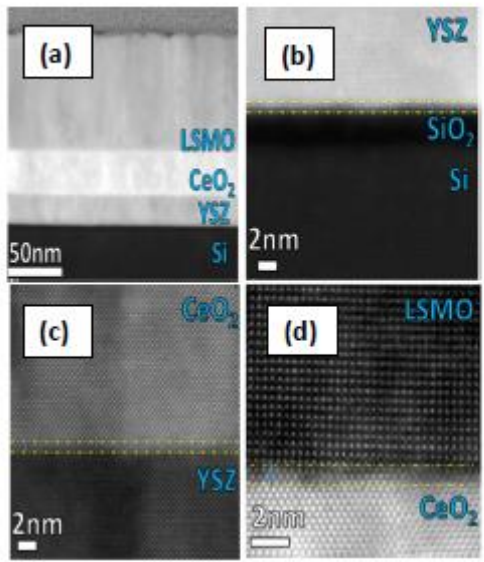

**Fig. 3**

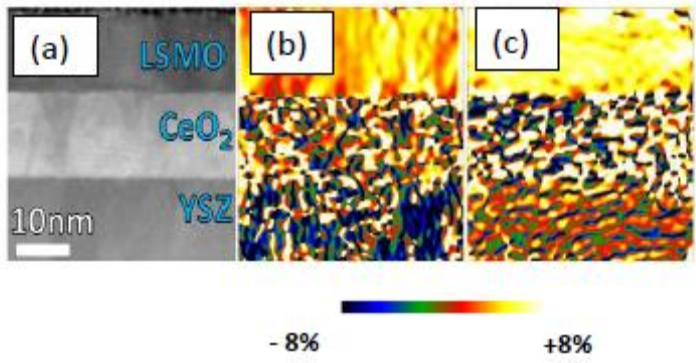

**Fig. 4**

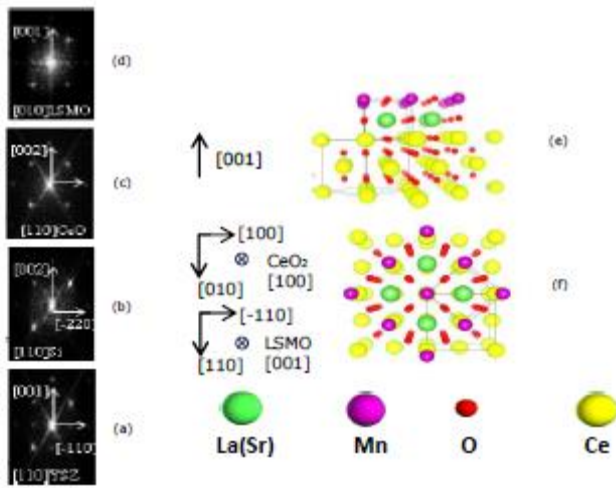

**Fig. 5**

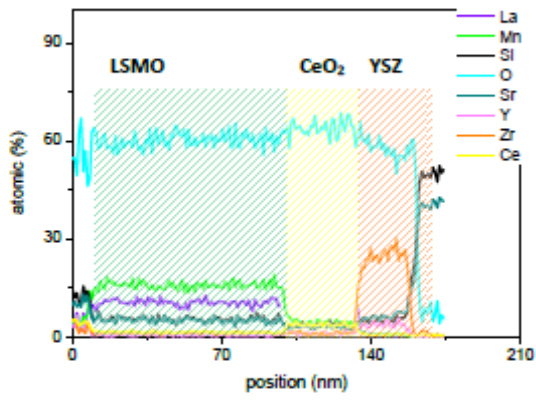

Fig. 6

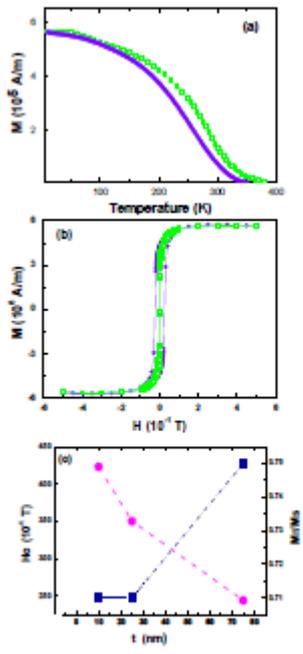

Fig. 7

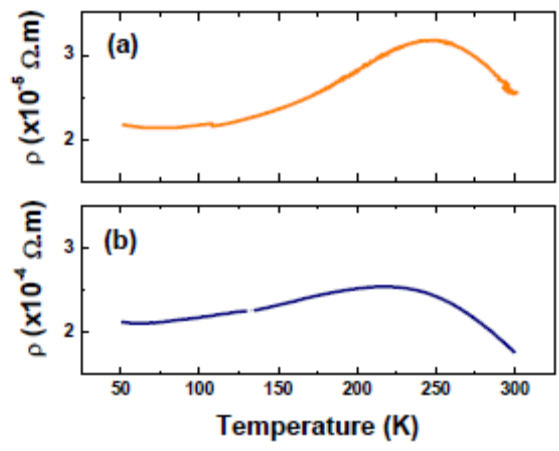

Fig. 8